\documentclass[letterpaper,twocolumn,prb]{revtex4}

\usepackage{amsmath,amsfonts,amssymb}
\usepackage[dvips]{graphicx}
\usepackage{enumerate}

\newcommand{\pt}[2]{\ensuremath{#1 \times 10^{#2}}}
\newcommand{\etal}{\emph{et al.}}
\newcommand{\etrans}{$^1\!\mathrm{B}_{2u}\!\!\leftarrow
  ^1\!\!\!\!\mathrm{A}_{1g}$}
\newcommand{\elexc}{$S_1$}
\newcommand{\vtrans}{$6_0^1$}
\newcommand{\transO}{$0^0$}
\newcommand{\transA}{A$_0^0$}
\newcommand{\transB}{A$_1^0$}
\newcommand{\transN}{A$_n^0$}
\newcommand{\vibration}[1]{$\nu_{#1}$}
\newcommand{\benzA}{C$_6$H$_6$}
\newcommand{\benzB}{C$_6$H$_5$D}
\newcommand{\benzC}{$s$-C$_6$H$_3$D$_3$}
\newcommand{\benzD}{C$_6$D$_6$}
\newcommand{\benzAd}{(\benzA)$_2$}
\newcommand{\benzBd}{(\benzB)$_2$}
\newcommand{\benzCd}{(\benzC)$_2$}
\newcommand{\benzDd}{(\benzD)$_2$}
\newcommand{\wn}{cm$^{-1}$}

\newcommand{\symm}[2]{\ensuremath{\text{#1}_{#2}}}

\newcommand{\mygraph}[1]{\includegraphics[scale=0.8]{#1}}
\newcommand{\molec}[1]{\includegraphics[scale=0.13]{#1}}
 
\begin{document}
 
\title{UV spectra of benzene isotopomers and dimers in helium
  nanodroplets}
\date{\today}
\author{Roman~Schmied}
\email{rschmied@princeton.edu}
\author{Pierre~\c Car\c cabal}
\altaffiliation[Current address: ]
  {Laboratory of Physical and Theoretical Chemistry,
   Oxford University,
   Oxford OX1 3QZ, United Kingdom.}
\author{Adriaan~M.~Dokter}
\altaffiliation[Current address: ]
  {Institute for Atomic and Molecular Physics,
   Kruislaan 407,
   Amsterdam, The Netherlands.}
\author{Vincent~P.~A.~Lonij}
\author{Kevin~K.~Lehmann}
\author{Giacinto~Scoles}
\affiliation{Department of Chemistry, Princeton University,
  Princeton, NJ 08544, U.S.A.}
 
\begin{abstract}
  We report spectra of various benzene isotopomers and their dimers in
  helium nanodroplets in the region of the first Herzberg-Teller
  allowed vibronic transition \vtrans\ \etrans\ (the \transA\ 
  transition) at $\sim$260\,nm. Excitation spectra have been recorded
  using both beam depletion detection and laser-induced fluorescence.
  Unlike for many larger aromatic molecules, the monomer spectra
  consist of a single ``zero-phonon'' line, blueshifted by
  $\sim$30\,\wn\ from the gas phase position.  Rotational band
  simulations show that the moments of inertia of \benzA\ in the
  nanodroplets are at least 6 times larger than in the gas phase.  The
  dimer spectra present the same vibronic fine structure (though
  modestly compressed) as previously observed in the gas phase. The
  fluorescence lifetime and quantum yield of the dimer are found to be
  equal to those of the monomer, implying substantial inhibition of
  excimer formation in the dimer in helium.
\end{abstract}
 
\maketitle

\section{Introduction}

Among all molecules, benzene occupies a very special place: it is an
organic molecule, with its conjugated ring forming the basis for a
vast field of chemistry; it is also effectively a very small molecule,
partially owing to its high degree of symmetry, and can thus be
studied with very high-resolution techniques, both
theoretical\cite{Spirko1999} and
experimental.\cite{Beck1979,Riedle1981,Oldani1984,Riedle1989,Okruss1999}
In fact, the benzene monomer and its isotopomers have been extensively
studied since the 1960s, when Callomon \etal\cite{Callomon1966} first
published a detailed analysis of moderate-resolution UV absorption
spectra of room temperature benzene vapor.  However, benzene van der
Waals interactions with itself and other molecules,\cite{Sun1996}
which are relevant to both biochemistry and molecular electronics,
have proved more demanding because of the nonrigid nature of most
complexes formed with benzene.  In particular, the structure and
internal degrees of freedom of the benzene dimer are still not firmly
established.

In the present work we have studied benzene and its dimer solvated in
superfluid helium nanodroplets, in order to further our understanding
of these systems.  Much is known about the gentle solvation effects in
helium nanodroplets,\cite{Toennies1998,dropletJCP2001} which allows us
to interpret the obtained spectra in great detail. In particular, two
molecules picked up by the same helium nanodroplet will rapidly
($\lesssim 1$\,ns) form a very cold van der Waals dimer, shedding
their relative kinetic and potential energy into the droplet and
reaching an equilibrium rotational and vibrational temperature of
$\sim$0.38\,K. In the past, helium nanodroplet isolation (HENDI) has
revealed rotationally resolved
spectra\cite{Hartmann1995,Callegari2001} of many molecules, which has
been interpreted as a demonstration of the superfluid nature of these
droplets. Electronic spectra have been recorded for a series of
molecules,\cite{Stienkemeier2001} in particular for higher polyacenes
(see Table~\ref{tb:shifts}) and other aromatic molecules. In the case
of glyoxal, the observation of a characteristic gap between the
zero-phonon line (ZPL) and the phonon wing provided the first direct
evidence for superfluidity in helium nanodroplets.\cite{Hartmann1996a}
The same gap was also observed for Na$_2$ molecules located on the
surface of the droplets.\cite{Higgins1998} For many other molecules,
the electronic spectra feature several ``zero-phonon'' lines, whose
origin has been extensively discussed in the
literature.\cite{Stienkemeier2001,Hartmann2001,Hartmann2002,Lehnig2003,Wewer2004,Lehnig2004}
While benzene has not been experimentally studied in helium droplets
before, several theoretical articles have appeared on this
subject.\cite{Kwon2001,Huang2003a,Patel2003}

We have focused our studies on the \transA\ vibronic transition of
benzene, which is the lowest Herzberg--Teller allowed transition to
the first electronic excited spin-singlet state (S$_1$, a $\pi
\rightarrow \pi^*$ excitation). It occurs around 259\,nm, and has a
long history of gas phase studies.\cite{Callomon1966} The S$_1$
$\leftarrow$ S$_0$ transition is symmetry forbidden, but
simultaneously exciting one quantum of the asymmetric ring breathing
vibrational mode \vibration{6} (521.4\,\wn\ for \benzA\ in the gas
phase)\cite{Atkinson1978c} breaks the 6-fold symmetry
(``Herzberg--Teller coupling'') and makes this vibronic transition
allowed by perpendicular-band one-photon absorption.  Higher
transitions in the \transN\ sequence include simultaneous excitation
of $n$ quanta of the totally symmetric \vibration{1} vibration
(923.538\,\wn\ for \benzA\ in the gas phase, see
Table~\ref{tb:lines}). The absorption oscillator strengths of the
\transN\ transitions are on the order of only \pt{1}{-4}
(Refs.~\onlinecite{Hiraya1991,Borges2003}), which has posed a
significant experimental challenge.

The isotopomers examined in this study were \benzA, \benzB, \benzC,
and \benzD, and their homo-dimers. Excitation spectra were recorded
both in helium droplet beam depletion using bolometric beam flux
detection, and in laser-induced fluorescence excitation. The former
method, ideally suited for poorly fluorescent
species,\cite{Carcabal2004} has given a better signal-to-noise ratio,
on the order of 100 or less; the latter method was not as successful
due to the relatively low fluorescence yield of benzene
($\sim$20\%).\cite{Spears1971} It was sufficient, however, for
measuring the fluorescence lifetimes of \benzA\ and \benzAd\ in helium
nanodroplets.

\section{Experimental}
\label{sec:experimental}

The apparatus is described in detail
elsewhere.\cite{Callegari2000,Callegari2001} In brief, helium droplets
are produced in a supersonic expansion at 59\,bar of research grade
$^4$He (99.9999\%)\cite{Spectra} through a 10\,$\mu$m nozzle at 14\,K,
and pass through a 390\,$\mu$m skimmer into the experimental chamber.
Under these conditions, the droplets have an average size of about
35\,000 atoms (extrapolating from the data of
Ref.~\onlinecite{Harms1998}).  Bolometric intensity measurements with
a chopped cluster beam yield $\sim$\pt{2.5}{14} helium atoms striking
the detector per second. Assuming 50\% clusterization and the above
mean cluster size, we estimate a total flux of $\sim$\pt{4}{9} helium
nanodroplets per second.  The droplets are doped while crossing a
2\,cm pickup cell, in which benzene is present at a pressure, on the
order of $10^{-4}$\,mbar, that can be optimized for a single or double
pickup. After doping they interact with a frequency-tripled kHz pulsed
Ti:Al$_2$O$_3$ laser in a wedged dielectric mirror multipass (98\%
reflective) with about 40 beam crossings.  A high repetition rate
laser, combined with a multipass cell, provides an acceptable duty
cycle of about 5\% cluster beam illumination, with the advantage of
easy access to higher harmonics.  The beam flux is detected with a
silicon bolometer attached to a $3\times3$\,mm$^2$ sapphire
slab,\cite{bolometer} with a noise-equivalent power of
$0.13\,\text{pW}/\sqrt{\text{Hz}}$. A collimator measuring 4\,mm in
diameter limits the cluster beam access to the sapphire slab. The
distance between the skimmer and the bolometer is $\sim$35\,cm.
Evaporation of the helium droplets following resonant excitation of
benzene molecules is measured by chopping the pulsed laser beam at the
fourth subharmonic of the laser repetition rate and amplifying the
bolometer signal through a cold J230 JFET,\cite{bolometer} a Stanford
SR550 preamplifier, and a SR510 lock-in amplifier. Detection at the
repetition rate of the laser could not be used due to the relatively
slow response of the bolometer and the high minimum repetition rate of
the pump laser.

The Ti:Al$_2$O$_3$ laser is a prototype Indigo\cite{PosLight} system
with a Fox--Smith resonator,\cite{Binks1997} pumped by an
Evolution-30\cite{PosLight} diode-pumped intracavity-doubled Nd:YLF
laser. It provides relatively narrow-band ($\sim$0.1\,\wn\ FWHM)
infrared pulses of $\sim$10\,ns duration at up to 3\,kHz repetition
rate.  Second-harmonic light was generated in an LBO crystal, and
combined with the fundamental in a BBO crystal to produce
third-harmonic radiation in the 260\,nm region, with an estimated
linewidth of 0.2\,\wn\ FWHM. The wavelength of the Ti:Al$_2$O$_3$
fundamental was measured with a Burleigh WA-4500 pulsed
wavemeter,\cite{Burleigh} and a Fe--Ne hollow-cathode lamp provided
absolute calibration at the second harmonic. We estimate our
wavelength calibration to be accurate to 0.02\,\wn\ and precise to
0.01\,\wn. For all measurements, the laser was running at a repetition
rate of 1083\,Hz, with a typical UV pulse energy of 30\,$\mu$J.

Fluorescence excitation spectra were acquired with a
Hamamatsu\cite{Hamamatsu} H5783-04 photomultiplier tube (PMT), which
has a cathode radiant sensitivity of about 40\,mA/W at 250\,nm,
corresponding to a quantum efficiency of $\sim$20\,\%. The PMT was
mounted at $90^{\circ}$ from the cluster and laser beams, and on the
order of 50\,\% of the fluorescence photons were collected with a
hemispheric/ellipsoidal aluminum mirror assembly.\cite{Hefter1988} The
gain of the PMT was estimated at $\sim$\pt{2}{6} by single-photon
observation on a fast oscilloscope. After dropping the collector
current over a 50\,$\Omega$ terminator, the PMT output was integrated
for 20\,ns (for lifetime measurements) or 50\,ns (spectra) with a
SR250 boxcar integrator, and the result fed into a SR510 lock-in
amplifier locked to a laser beam chopper running at 1/4 of the laser
repetition rate. The laser beam was chopped to eliminate electrical
pickup from firing of the laser Q-switch.

All spectra were measured with 1\,Hz bandwidth on the lock-in
amplifier, and recorded through a computer interface with 1\,Hz
sampling rate.  Typical scanning speeds were on the order of
1\,\wn/min. The resulting data (pairs of wavelength and signal
amplitude) were Gaussian smoothed with a standard deviation of
0.1\,\wn.

In all isotopomers of benzene we used, carbon isotopes were present at
natural abundances.  \benzB\ (98+\%) and \benzC\ 
(\emph{sym}-benzene-d$_3$, 1,3,5-C$_6$H$_3$D$_3$, 98\%) were purchased
from Aldrich;\cite{Aldrich} \benzD\ (99.5\%) was purchased from
Cambridge Isotope Laboratories.\cite{Cambridge} All isotopomers were
used without further purification. A $^1$H NMR spectrum of a
\benzA/\benzD\ mixture showed no significant impurities.

\section{Beam Depletion Spectra}

The recorded beam depletion spectra were assigned by studying line
intensities as functions of the dopant pressure in the pickup cell,
verifying the typical Poisson distribution expected for uncorrelated
pickup events.\cite{Lewerenz1995} We will present and discuss the
spectra in five sections, separately considering the line shapes of
the monomer, those of the dimer, the complexes with argon, the droplet
solvation shifts, and the excitation of a totally symmetric
vibrational mode.

\subsection{Monomer Spectra}
\label{sec:lineshape}

The \transA\ absorption lines of all the benzene monomer isotopomers
we have studied were found to present the same vibronic fine structure
as in the gas phase, consisting of a single absorption line within our
experimental resolution (two lines in the case of \benzB, see
Fig.~\ref{fig:monomerlines}).  Figure~\ref{fig:C6H6_monomer_A00} shows
the \transA\ beam depletion spectrum of \benzA\ with a signal-to-noise
ratio of $\sim$100; the linewidth is 0.53\,\wn\ full width at half
maximum (FWHM) and the peak is at 38636.47\,\wn. No phonon wing was
detected within 15\,\wn\ to the blue of the transition, and we
estimate that at least 80\% of the spectral intensity is in the
zero-phonon line. It has been shown for a variety of molecules that
droplet phonon wings are often weak and require significant saturation
of the ZPL in order to be evident (see
Ref.~\onlinecite{Stienkemeier2001} for a review). As our conditions
are far from saturation (see Section~\ref{sec:amplitude}), and the
$\pi \rightarrow \pi^*$ valence excitation of benzene is relatively
weakly coupled to the first helium solvation shell (see
Section~\ref{sec:shifts}), our inability to detect a phonon wing can
be rationalized.

Diffusion Monte-Carlo (DMC) studies of the \benzA--He$_{14}$ system
have predicted a multitude of excited states of the first helium
solvation shell with energies of 10 to 20\,\wn\ 
(Ref.~\onlinecite{Huang2003a}). However, UV transitions to helium
excited states as computed in Ref.~\onlinecite{Huang2003a} are
symmetry-forbidden: helium excitations must be of \symm{A}{1g}
symmetry to be excited in the \transA\ transition,\cite{SymmetryNote}
which is mutually exclusive with the POITSE (projection operator,
imaginary time spectral evolution) method of
Ref.~\onlinecite{Huang2003a}.  Therefore, any helium states computed
in Ref.~\onlinecite{Huang2003a} are invisible in direct UV
spectroscopy, and any \symm{A}{1g} phonons that we could potentially
detect are not computable with POITSE.  However, the energy of the
$(L,M)=(2,0)$ phonon, which can in principle be excited in the
\transA\ transition, is unlikely to be very different from that of the
other $L=2$ phonons computed in Ref.~\onlinecite{Huang2003a}
($\sim$12\,$k_{\text{B}}$K), judging from the small $M$-dependence of
the energies of the POITSE phonons in \benzA--He$_{14}$. The energy of
the $(L,M)=(0,0)$ phonon, on the other hand, cannot be inferred easily
because it involves a qualitatively different helium motion, being the
only phonon that varies the mean density of the helium in the first
solvation shell.  Pending a calculation of energies and Franck--Condon
factors for these excitations, we conclude that the lack of phonon
wings in our spectra are due to insufficient signal-to-noise ratios.

A comparison of the \benzA\ \transA\ droplet spectrum to a simulated
gas phase spectrum at 0.38\,K (the experimental rotational temperature
of molecules in helium droplets),\cite{Hartmann1995,Callegari2001}
using the Hamiltonian of Ref.~\onlinecite{Okruss1999} (up to
$J''=20$), shows that the droplet spectrum is significantly
compressed, suggesting that the effective moments of inertia of
benzene are much larger in the droplet than in the gas phase (see
Figure~\ref{fig:comparespectra}).  As shown in
Fig.~\ref{fig:stickspectrum}, the droplet spectrum can be approximated
by introducing two scaling factors $\kappa_B$ (scaling $B'$ and $B''$)
and $\kappa_C$ (scaling $C'$, $C_0'\zeta'$, and $C''$) that describe
the effect of the helium droplet on the rotational motion of the
benzene molecules. Smaller values of $\kappa_B$ and $\kappa_C$ result
in narrower spectra, with less recognizable structure due to the
Gaussian smoothing involved; since our spectrum consists of only one
peak, any values of $\kappa_B$ and $\kappa_C$ extracted from line fits
can only be interpreted as upper limits. In this sense, we estimate
that both $\kappa_B$ and $\kappa_C$ are smaller than $\sim$1/6.  This
means that the effective moments of inertia of \benzA\ in superfluid
helium are at least 6 times larger than in the gas phase.  While a
rigid model of the first solvation shell agrees with this limit on
$\kappa_B$, it overestimates $\kappa_C$: a rigid \benzA--He$_{14}$
cluster, with two helium atoms on the $C_6$ axis 3.3\,\AA\ from the
origin and 12 helium atoms in two rings of 6 atoms each located
3.9\,\AA\ from the origin at an angle of 45$^{\circ}$ from the $C_6$
axis, estimates $\kappa_B^{\text{rigid}} \approx 0.12$ and
$\kappa_C^{\text{rigid}} \approx 0.33$.  Note that our upper limits on
$\kappa_B$ and $\kappa_C$ are still only about half the size of the
typical values of $\sim$$1/3$ for large molecules.\cite{Callegari2001}
The optimized widths of the smoothing functions used to produce the
fitting spectra of Figure~\ref{fig:stickspectrum}, on the order of
0.4\,\wn, are significantly larger than the laser linewidth
($\sim$0.2\,\wn). Many electronic spectra of molecules in helium
nanodroplets have been recorded with similar
linewidths,\cite{Stienkemeier2001} while purely vibrational HENDI
spectra routinely achieve a hundred times smaller
linewidths.\cite{Callegari2001} In the case of benzene, the observed
line profiles are close to Gaussian, with little asymmetry (see
Figure~\ref{fig:C6H6_monomer_A00}). Rapid vibrational relaxation in
\elexc\ with a lifetime on the order of 10\,ps could account for the
observed linewidth, but it would lead to more pronounced Lorentzian
line wings. Inhomogeneous broadening due to either the cluster size
distribution\cite{Dick2001} or to translational levels of the benzene
molecules in the clusters\cite{Lehmann1999b} is expected to result in
asymmetric lines.

\subsection{Dimer spectra}
\label{sec:dimersplit}

Figure~\ref{fig:dimerlines} shows the \transA\ beam depletion spectra
of all benzene homo-dimers considered in this study. The linewidths of
all sharp features are $\sim$0.5\,\wn, very similar to the 0.53\,\wn\ 
linewidth of the \benzA\ \transA\ line (see
Section~\ref{sec:lineshape}).

For the benzene dimer in the gas phase, not only the \transN\ vibronic
progression can be observed, but the reduced symmetry in the T-shaped
dimer (see below) makes also the electronic origin transition \transO\ 
very weakly allowed.  For \benzAd\ in the gas phase, both the \transO\ 
and \transA\ UV transitions are split, featuring two peaks of similar
linewidth and an intensity ratio around 1:1.4 (see
Refs.~\onlinecite{Law1984,Boernsen1986}). The \transA\ transition also
features a weaker progression of modes spaced by about 17\,\wn, which
has been assigned to the stretching of the van der Waals dimer
bond.\cite{Scherzer1992} While we have observed neither the \transO\ 
transition nor the van der Waals progression in helium droplets, the
splitting of the \transA\ transition is present in helium droplets
with the same intensity ratio as in the gas phase (see lowest graph in
Fig.~\ref{fig:dimerlines} and Fig.~\ref{fig:lineratio}), though
modestly compressed. In \benzDd, the \transO\ and \transA\ transitions
were found in the gas phase to consist of a sharp line with a broader,
weaker peak further to the blue,\cite{Law1984,Boernsen1986} a pattern
which again was reproduced in helium droplets.  Apart from a 20\%
smaller spacing for the two \transA\ lines in \benzAd\ (3.7\,\wn\ in
the gas phase, 2.90\,\wn\ in helium droplets), we conclude that the
mechanism leading to the splitting is mostly unperturbed by the helium
droplet.  This observation must be reconciled with the large increase
in the effective moments of inertia of the individual benzene
molecules in helium (Section~\ref{sec:lineshape}).

We will now try to establish what mechanisms can or cannot lead to the
observed benzene dimer splitting; we will lead this discussion in a
similar manner to that of Ref.~\onlinecite{Sun1996}, where the same
issues have been discussed for the benzene dimer in the gas phase.

It was established in the gas phase work that the dimer splitting is
not due to two different conformers with overlapping spectra, since
the relative intensities do not depend on source
conditions.\cite{Henson1991} Further, the efficient relaxation and
long-range quadrupole-quadrupole alignment\cite{Nauta1999} in helium
droplets, which can align the two benzene monomers over a distance of
about 30\,\AA, would most probably lead to intensity ratios different
from those seen in jet spectra. More convincingly, hole-burning
studies\cite{Scherzer1992} have been able to identify two weak
secondary conformers of \benzAd, none of which is the source of the
dimer splitting and both of which are similarly split. We have not
been able to detect these conformers in helium droplets, as expected
due to their smaller intensities.

Microwave spectra\cite{Arunan1993} have determined that the benzene
dimer is T-shaped, with a fast internal rotation around the dimer axis
which leads to a symmetric-top rotational spectrum. Further arguments
that favor a T-shaped over a displaced-parallel dimer are (i) the
perpendicular orientation of nearest neighbors in solid
benzene,\cite{Cox1958,Bernstein1969} (ii) a permanent electric dipole
in at least one conformer of the benzene
dimer,\cite{Janda1975,Arunan1993} (iii) the demonstration that the two
benzene molecules occupy inequivalent sites in the
dimer,\cite{Henson1991,Henson1992} and (iv) recent
CCSD(T)\cite{Spirko1999} and AIMI\cite{Tsusuki2002} ab-initio
calculations.  It is thus natural to suggest that the splitting is due
to absorption in the two inequivalent monomers.  However, studies of
\benzA--\benzD\ hetero-dimers have
shown\cite{Law1984,Scherzer1992,Henson1992,Venturo1993} that in the
main \transA\ transition lines, the ``stem'' monomer is almost
exclusively excited, with a characteristically split spectrum, whereas
the spectrum of the ``top'' monomer consists of an extended
($\sim$100\,\wn) van der Waals progression with peak intensities less
than 10\% of those of the ``stem'' monomer.  Therefore, the existence
of inequivalent monomers cannot account directly for the dimer
splitting.

The reduced symmetry of the ``stem'' monomer can lift the degeneracy
of E-type vibrations.\cite{Hopkins1981} In fact, the molecules in
solid benzene do not have sixfold symmetry and their degenerate
vibrational modes are split by an amount very similar to the dimer
splitting.\cite{Bernstein1969} While a reduced symmetry would split
the \transA\ transition (\vibration{6} is an \symm{E}{2g} vibration),
it cannot split the non-degenerate \transO\ transition, as observed in
the gas phase. Unless the splittings of these two transitions have
different origins (which is unlikely because of their qualitative
similarity) we can exclude symmetry breaking as a cause for the dimer
splitting.  Moreover, the fluorescence lifetime of the benzene dimer
in helium droplets is found to be equal to that of the gas phase
monomer (Section~\ref{sec:fluorescence}), which would be unlikely if
the ``stem'' monomer was significantly distorted since this would
decrease its fluorescence lifetime. Note that a first-order
perturbation of a degenerate vibronic level will result in a line
doublet of equal intensities, as in the case of \benzB\ (see
Fig.~\ref{fig:monomerlines}); the very asymmetric intensity patterns
observed in the dimers, in particular for \benzDd, cannot be easily
explained with such a symmetry-breaking mechanism.

Excitonic interactions\cite{Boernsen1986,Henson1992} between the two
benzene monomers can be ruled out as the source of the dimer
splitting: the \transA\ spectrum of the \benzA--\benzD\ 
hetero-dimer\cite{Law1984,Scherzer1992} has been found to consist
mainly of two line doublets, one doublet very similiar to the \benzAd\ 
spectrum and the other similar to the \benzDd\ spectrum. If the
splitting was due to exciton hopping, it would be absent in
hetero-dimers, where excitonic interactions are non-resonant.

It is interesting to note that the reduced $D_{3h}$ symmetry of
\benzC\ still allows for $E$-type symmetry species, resulting in a
single \transA\ absorption line (see Fig.~\ref{fig:monomerlines}).
This signifies that, for most purposes related to the experiments
reported here, the \benzC\ molecule can be regarded as standing
halfway between \benzA\ and \benzD. The sequence of \benzAd, \benzCd,
and \benzDd\ \transA\ spectra show a decreasing splitting and an
increasing line intensity ratio. The regularity of this progression
suggests that the masses, and not the particular nuclear spin weights,
of the isotopomers determine their spectra.

While tunneling of the various intermolecular motions in the dimer
(librations and interchange) can lead to splittings, these are too
small in magnitude and in any case limited by the associated
rotational constant, which can be no larger than 0.28\,\wn\ (for
rotation around the inter-monomer axis).  A tentative assignment of
microwave spectra predicts small splittings of 15 or 30\,kHz due to
interconversion tunneling;\cite{Arunan1993} theoretical estimates of
various tunneling splittings are even smaller.\cite{Spirko1999}

The most probable explanation for the dimer splitting is librational
motion. A theoretical analysis\cite{Spirko1999} of the T-shaped dimer
predicts intermolecular vibrational frequencies as low as 3.4\,\wn\ 
for the torsion around the dimer axis (i), and 4.8\,\wn\ for the
``stem'' rotation around its $C_6$ axis perpendicular to the dimer
axis (ii); the interchange vibration (iii) is projected at 13.6\,\wn,
much larger than the observed splitting.  In all of these modes, the
helium solvation shell increases both the effective mass (see
Section~\ref{sec:lineshape}) and the curvature of the vibrational
potential energy surface (one or two helium atoms are expected to
localize on either side of the ``stem'' monomer,\cite{Kwon2001}
further attracted to the ``top'' monomer $\pi$-cloud, thus hindering
benzene dimer intermolecular motion). Sophisticated simulations,
accounting for the effects of helium solvation and muclear spin
statistical weights,\cite{Odutola1981,Schmied2004} are called for in
order to judge these assumptions.

There are two ways in which librational motion can lead to the
observed dimer splittings. Firstly, conservation of nuclear
spin\cite{Harms1997} will populate several internal-motion tunneling
levels even at very low temperatures, which will have different
spectra if there is a change in the intermolecular force constants
upon electronic excitation.  This has been conclusively ruled out as
the origin of the dimer splitting by hole-burning
studies\cite{Scherzer1992} of the \benzA--\benzD\ hetero-dimer.
Secondly, excitation of a varying number of librational quanta leads
to a series of absorption lines, which is what we believe is the
origin of the dimer splitting.

Microwave spectra suggest that in the gas phase, the ``stem'' rotation
around the dimer axis [mode (i)] has such a large zero-point motion
that the T-shaped benzene dimer effectively behaves as a symmetric
top; in helium droplets, however, this may not be the case since the
helium strongly suppresses rotational motion, as discussed in
Section~\ref{sec:lineshape}. A study of selection rules and transition
moments is required in order to determine which librational mode can
be excited in the \transA\ transition of the dimer.

The added complexity of the \benzBd\ and \benzCd\ spectra is most
likely due to the naturally occurring different conformers. In
\benzCd, the ``stem'' monomer can have a hydrogen or a deuterium atom
pointing toward the ``top'' monomer, resulting in two slightly
different conformers. Similarly, \benzBd\ has four conformers
depending on the orientation of the ``stem'' monomer.

\subsection{Complexes with Argon}
\label{sec:argon}

Apart from monomer and dimer spectra of benzene, we have observed the
\transA\ spectrum of the \benzA--Ar dimer: Figure~\ref{fig:overview}
shows a spectrum featuring \benzA, \benzA--Ar, and \benzAd\ 
absorptions. Despite several independent attempts, no further
complexes of benzene with argon were detected. In particular, we have
tried to detect \benzAd--Ar$_n$ and \benzA--Ar$_2$ complexes, in order
to gain structural information of the doped helium clusters.

In the gas phase, there are two conformers of \benzA--Ar$_2$, with
both argon atoms on the same side of the benzene plane (17\,\wn\ 
redshift)\cite{Schmidt1991a} or on opposite sides (41.869\,\wn\ 
redshift).\cite{Neusser1994} In helium droplets, as in the gas phase,
the latter and more stable\cite{Schmidt1991a} ``sandwich'' complex is
expected to coincide\cite{Weber1991b} with the bluer one of the two
\benzAd\ lines, since its redshift with respect to the \benzA\ 
absorption line is twice that of \benzA--Ar, assuming the two argon
atoms to be independent.  A study of the dependence of the dimer line
ratio on argon pressure in the second pickup cell (see
Fig.~\ref{fig:lineratio}) did not reveal any significant variation
even for argon pressures that are optimized for about two pickups
($\sim$\pt{16}{-4}\,mbar), where the signal of the \benzA--Ar$_2$
complex should have an amplitude comparable to that of \benzAd.  At
this point we are unable to explain the absence of these complexes.

\subsection{Solvation and interaction shifts}
\label{sec:shifts}

Table~\ref{tb:lines} lists experimental transition energies measured
in helium droplets, along with the corresponding gas phase values
wherever known. Previously, all dimer transition energies were given
with respect to inaccurate monomer lines, which have been determined
from contour fits of room-temperature spectra (see, \emph{e.g.},
Ref.~\onlinecite{Callomon1966}) and are off by $+2.4(2)$\,\wn\ for
\benzA\ and $+1.1(1)$\,\wn\ for \benzD. The dimer transition energies
listed in Table~\ref{tb:lines} have been computed from the more
accurate monomer transitions of
Refs.~\onlinecite{Okruss1999,Weber1991a}, assuming that the
dimerization shifts of Refs.~\onlinecite{Boernsen1986,Law1984} are
accurate.

Table~\ref{tb:shifts} summarizes experimental helium solvation shifts
of the smallest polyacenes, with an apparent tendency of increased
redshifts with increasing molecular size.  The \benzA\ blueshift of
$+30.31$\,\wn, as qualitatively predicted in
Ref.~\onlinecite{Even2001}, agrees with previous experimental data on
the \transA\ transition of the benzene--helium system: rotationally
resolved supersonic jet spectra of \benzA--He and \benzA--He$_2$, with
the helium atoms on the $C_6$ axis, have found blueshifts of
2.31(2)\,\wn\ per helium atom,\cite{Beck1979} and room-temperature
spectra of benzene in high-pressure helium gas\cite{Nowak1987} have
measured a $\sim$10\,\wn\ blueshift at the helium density
corresponding to that of bulk liquid helium (21.8\,nm$^{-3}$). These
shifts are a combination of (i) a small dispersive redshift, due to
the small polarizability of the solvating helium and the increased
polarizability of benzene in the excited state (see
Table~\ref{tb:moments}), only slightly counteracted by a decreased
quadrupole moment, and (ii) a larger electronic blueshift due to the
expansion of the electron cloud against the first helium solvation
shell upon excitation. In fact, the helium atoms in \benzA--He and
\benzA--He$_2$ recede from the benzene molecule by about 7--8\,pm upon
electronic excitation of benzene.\cite{Beck1979} Note, however, that
this expansion of the electron cloud is not reproduced in the second
moments of the benzene charge distribution (see
Table~\ref{tb:moments}), although the electronic excitation is more
than halfway to the ionization threshold
(74556.57\,\wn).\cite{Neuhauser1997}
The \vibration{6} vibrational contribution to the droplet shift is
probably much smaller than the electronic contribution, and comparable
to the \vibration{1} vibrational droplet shift of only $+0.72$\,\wn\ 
(see Section~\ref{sec:A10transition}).

The \benzAd\ dimerization redshift (the wavenumber difference between
the monomer and averaged dimer bands, both in helium droplets) is
decreased to 35.7\,\wn\ from the gas phase value of 42.3\,\wn\ 
(Ref.~\onlinecite{Boernsen1986}).  Similarly, the \benzA--Ar
complexation redshift is decreased to 17.20\,\wn\ from the gas phase
value of 21.087\,\wn\ (see Table~\ref{tb:lines}).  These results are
not easily explained: in both cases we would expect a \emph{larger}
complexation redshift in helium than in the gas phase, because (i) the
adducts replace helium atoms that contributed to the 30.31\,\wn\ 
blueshift of the benzene monomer, and (ii) the van der Waals bond
between the adduct and the benzene monomer tends to be weakened in
helium droplets with respect to the gas phase, reducing the repulsive
dimerization blueshift.  The latter point, suggested by a decreased
exciton splitting for (SF$_6$)$_2$ in helium
droplets,\cite{Hartmann1996b} is substantiated by a density-functional
(DF) calculation of the \benzA--Ar system in a helium droplet (500
helium atoms) assuming cylindrical symmetry (Ar located on the $C_6$
axis of \benzA), using the Orsay-Paris functional,\cite{DupontRoc1990}
with the CCSD(T) \benzA--He potential from Ref.~\onlinecite{Lee2003}
(averaged cylindrically) and the Ar--He potential from
Ref.~\onlinecite{Ahlrichs1977}.  At the \benzA--Ar equilibrium
distance\cite{Brupbacher1994} of 3.5\,\AA, the DF calculation
estimates a force of 1.8\,pN pulling the moieties apart, resulting in
a slight bond stretching of 0.5\,pm (assuming a harmonic force
constant of 3.5\,N/m).\cite{Brupbacher1994}

\subsection{\transB\ transition}
\label{sec:A10transition}

For a \benzA\ molecule in helium nanodroplets, the wavenumber of one
quantum of \vibration{1} vibration added to the \vibration{6} excited
state of \elexc\ is 924.26\,\wn, compared to 923.538\,\wn\ in the gas
phase (see Table~\ref{tb:lines}). Both \vibration{1} and \vibration{6}
are mostly vibrations of the carbon ring, with the hydrogen atoms
almost stationary;\cite{Pulay1981} the very small droplet blueshift of
only $0.72$\,\wn\ suggests that these modes are only weakly perturbed
by the helium droplet. Moreover, in the \benzAd\ dimer, the shift
between the center of the two main lines of the \transB\ spectrum and
the center of the \transA\ lines (see Fig.~\ref{fig:dimerlines}) is
924.32\,\wn, almost indistinguishable from the monomer value of
924.26\,\wn. As we excite mostly the ``stem'' monomer in these UV
transitions (see Section~\ref{sec:dimersplit}), the \vibration{1}
vibration is only slightly influenced by the presence of the ``top''
monomer.

The \benzAd\ dimer splitting, as discussed in
Section~\ref{sec:dimersplit}, is reduced to 2.52\,\wn\ in the \transB\ 
transition (see the lowest panel in Fig.~\ref{fig:dimerlines}).
Further, there is an additional, weaker absorption line similarly
spaced to the red. In the gas phase, there are no symmetry-allowed
transitions close to \transB\ (Ref.~\onlinecite{Atkinson1978a}).  This
observation is consistent with the proposed explanation of the dimer
splittings through librational excitations, assuming that an
additional librational excitation becomes accessible due to a
modification of the van der Waals interaction in the $\nu_6+\nu_1$
mode. If this assignment is correct, however, then the above
consideration about the \vibration{1} energy in \benzAd\ becomes
questionable, since we do not know which lines in the \transA\ 
spectrum correspond to which in the \transB\ spectrum.

\section{Fluorescence excitation spectra}
\label{sec:fluorescence}

The fluorescence excitation spectra yielded approximately the same
energies for the monomer and dimer transitions as the beam depletion
spectra (see Fig.~\ref{fig:C6H6_monomer_A00}).  The small differences
are line shape distortions due to fluctuations in the laser power,
since the signal-to-noise ratio of the fluorescence excitation spectra
was smaller; these differences are not significant for the following
discussion.

By sweeping the delay of the boxcar integrator with respect to the
laser pulse, using a fixed 20\,ns gate width, we determined the
fluorescence lifetime of the \benzA\ monomer to be 115(5)\,ns. This is
comparable to the 103\,ns previously reported in jet expansion
experiments.\cite{Spears1971,Hopkins1981} The value of 103\,ns is
actually for the electronic origin (\transO) excitation, whereas we
are exiting \transA\ (gas phase lifetime: 79\,ns).\cite{Spears1971}
However, the additional vibrational energy probably relaxes rapidly
into the helium droplet, and thus we are seeing fluorescence from the
\transO\ state.

For both components of the dimer \transA\ transition we find
120(10)\,ns lifetime, which is significantly longer than the gas phase
value of $\sim$40\,ns (Ref.~\onlinecite{Law1984}). This short gas
phase lifetime has been attributed to a mixed state of excited van der
Waals dimer and excimer\cite{Shinohara1989} or the rapid conversion
from the former to the latter upon excitation.\cite{Hopkins1981} We
therefore conclude that excimer formation of the benzene dimer is
suppressed in helium droplets.

\section{Signal Amplitudes}
\label{sec:amplitude}

The oscillator strengths $f$ of the \transN\ transitions in \benzA\ 
are on the order of \pt{0.5-2}{-4}
(Refs.~\onlinecite{Hiraya1991,Borges2003}). Using the measured
linewidth of 0.53\,\wn\ for the \transA\ transition (see
Fig.~\ref{fig:C6H6_monomer_A00}) and $f=\pt{1.37}{-4}$
(Ref.~\onlinecite{Hiraya1991}), the peak absorption cross section is
about 0.7\,\AA$^2$. A typical 30\,$\mu$J UV laser pulse with 2\,mm
beam diameter thus excites about 9\% of the benzene molecules in its
field. In a multipass cell with 98\% reflectivity, 40 beam crossings
thus excite a number of benzene molecules that is equal to the number
of doped clusters in about 5\,mm of the cluster beam (assuming no
saturation effects).

The rate at which benzene molecules are excited is $\eta S/(h \nu)$,
where $S$ is the lock-in signal (converted from Volts to Watts through
the bolometer sensitivity \pt{2.0}{5}\,V/W), $h \nu$ is the photon
energy, and $\eta\approx 0.2$ is the ratio of the energy it takes to
evaporate a helium atom from a droplet
($\sim$7\,$k_{\text{B}}$K)\cite{Chin1995} to the energy that this
process removes from the beam flux (the kinetic energy $\frac52
k_{\text{B}}T_{\text{nozzle}}=35$\,$k_{\text{B}}$K).  Complete
accommodation of the cluster beam on the bolometer is assumed, which
may not be accurate for our ``warm'' bolometer operating at 4--5\,K.
At a wavelength of 259\,nm, a typical peak lock-in signal level of
5\,pW thus corresponds to $\sim$\pt{1.3}{6} excited benzene molecules
per second, or $\sim$\pt{2.4}{3} per laser pulse. This suggests that
there are $\sim$\pt{2.4}{3} benzene-doped helium droplets in 5\,mm of
cluster beam (see above); since the beam moves at $\sim$380\,m/s, the
total flux of benzene-doped helium clusters is estimated to be
$\sim$\pt{2}{8}\,s$^{-1}$, or about 5\% of the total number of
clusters (see Section~\ref{sec:experimental}), compared to $1/e$
(37\%) predicted for optimal single pickup. This mismatch could be due
to either an inaccurate knowledge of the mean cluster size, or the
inaccuracy of the assumption of 50\% clusterization of the helium
beam.

Under the above assumptions, the first crossing of the laser with the
cluster beam excites $\sim$90 benzene molecules per shot.  On the
other hand, for the fluorescence signal in
Fig.~\ref{fig:C6H6_monomer_A00} we estimate the production of about
0.5 primary electrons in the PMT per laser shot.
For this spectrum, we used a 50\,ns gate delayed by about 20\,ns from
the UV laser pulse; this delay was necessary to avoid light scattered
by the droplet beam, which was about twice as intense as the
fluorescence signal. With a fluorescence lifetime of $\sim$115\,ns
(see Section~\ref{sec:fluorescence}), we thus captured about 30\% of
the total fluorescence signal. Compounding this with the 20\% quantum
efficiency of the PMT and 50\% collection efficiency, we estimate
$\sim$20 fluorescence events per laser shot.

These signal estimates yield on the order of four to five times less
fluorescence events than relaxation/evaporation events per crossing of
droplet and laser beams, suggesting a fluorescence quantum yield of
20\%, very close to the experimental fluorescence yield of 20\% in the
gas phase.\cite{Spears1971} This demonstrates that comparison of
photon emission and beam depletion signals can be used to estimate
absolute quantum yields of emission.

For the benzene dimer, the peak beam depletion signal is about half
the peak monomer signal (see Figure~\ref{fig:dimerlines}), while the
fluorescence signals of monomers and dimers were found to be of
similar amplitude. This indicates a larger fluorescence quantum yield
for \benzAd\ than for \benzA\ in helium droplets

\section{Conclusions}

Helium nanodroplet isolation electronic spectroscopy of aromatic
molecules has been extended to benzene and its dimer.
Figure~\ref{fig:overview} shows a typical broad scan of several types
of dopants in helium droplets. While the cluster beam depletion
spectra are very similar to gas phase spectra, they are blueshifted by
about 30\,\wn, confirming an apparent trend\cite{Stienkemeier2001} of
larger blueshifts for larger excitation energies in aromatic molecules
(see Table~\ref{tb:shifts}). The solvation blueshift in benzene is
mostly electronic, judging from the \vibration{1} vibration which was
found to be blueshifted by only 0.72\,\wn\ in helium droplets.

The coupling of the electronic excitation in benzene to the
surrounding helium droplet does not lead to any observable signature,
and we estimate that at least 80\% of the electronic absorption
intensity is in the zero-phonon line.

Not surprisingly, the effective rotational moments of inertia of
\benzA\ were found to be greatly increased due to the surrounding
helium. While rotational transitions could not be resolved, a contour
fit estimates that the moments of inertia are larger by a factor of
\emph{at least} 6.

The splitting patterns of the \transA\ transition in \benzAd\ and
\benzDd\ are modestly compressed with respect to the gas phase, but
not qualitatively altered; as their most plausible source we find the
excitation of librational intermolecular motion, although the details
remain unknown.  Though the spectra of \benzAd\ and \benzDd\ are
substantially different, the \transA\ spectrum of \benzCd\ appears to
be intermediate between the two, indicating that the dimer splitting
differences are related only to the isotopic masses and not to their
nuclear spin weights.

The fluorescence quantum yield of \benzA\ in helium nanodroplets
agrees with the gas phase value. The fluorescence lifetime of \benzAd,
however, is identical to that of \benzA\ in helium droplets, in stark
contrast to the much reduced lifetime of \benzAd\ in the gas phase.
Further, the fluorescence quantum yield of the dimer was not lower
than that of the monomer, contrary to gas phase
studies\cite{Spears1971,Hopkins1981} that found these to differ by
about an order of magnitude.  Both the short lifetime and low
fluorescence yield of the gas phase dimer have been attributed to
excimer formation upon electronic excitation. In the light of our
observation of sharply increased moments of inertia of benzene in
helium (Section~\ref{sec:lineshape}), it is likely that the
transformation from the T-shaped van der Waals dimer to the excimer,
which is believed to have a parallel stacked
configuration,\cite{LangridgeSmith1981,Hopkins1981} is inhibited by
the presence of helium.  The fluorescence lifetime we observe would
then be the lifetime of the pure van der Waals dimer, expected to be
similar or slightly shorter (because of an increased internal
conversion rate due to the weak van der Waals bond) than that of the
monomer. We would in fact expect it to be similar to the lifetime of
the trimer (81\,ns, Ref.~\onlinecite{Hopkins1981}; 79\,ns,
Ref.~\onlinecite{Hirata1999}), which does not form an excimer.

\acknowledgments

The authors would like to thank Marcel Nooijen for valuable
discussions about electronic excited state properties, and Carlos
Pacheco for the NMR spectra. This work was supported by the National
Science Foundation.
 

\clearpage

\bibliography{benzene}

\begin{thebibliography}{84}
\bibitem[{Sym()}]{SymmetryNote}
\bibinfo{note}{More precisely, the phonons must have \symm{A}{1g} or
  \symm{A}{2g} symmetry to be excited in the \transA\ transition. The
  \symm{A}{1g} symmetry corresponds to phonons with even angular momentum $L$
  and projection $|M|=0,6,12,\dotsc$ onto the $z$ axis defined by the benzene
  molecule $C_6$ axis; the \symm{A}{2g} symmetry corresponds to even $L$ and
  $|M|=6,12,\dotsc$. Assuming that excitation of phonons with $|M|\ge 6$ is
  weak, only phonons of \symm{A}{1g} symmetry need to be considered.}
\end{thebibliography}

\clearpage

\begin{table}
  \centering
  \caption{Absorption line positions of various benzene isotopomers and
  dimers in helium nanodroplets (beam depletion spectra). Units are
  wavenumbers (\wn); uncertainties in present measurements are
  0.02\,\wn\ where not specified.  Measurements are peak positions of
  least-squares fitted Gaussians. The measurement in square brackets
  refers to the band origin of a fitted spectrum (see
  Figure~\ref{fig:stickspectrum}).
  Energies labeled with an asterisk have been adjusted with more
  accurate values of the respective monomer transitions (see text).}
  \label{tb:lines}
  \begin{ruledtabular}
    \scriptsize
    \begin{tabular}{llllcl}
      molecule & line & droplet & gas phase & Ref. & shift\\
      \hline
      \benzA & \transA & 38636.47 & 38606.098(2) & \onlinecite{Okruss1999} & $+30.37$\\
      & & $[38636.41]$ & & & $[+30.31]$\\
      \benzB & \transA$_{\text{(a)}}$ & 38664.85 & 38634.2429(1) & \onlinecite{Riedle1994} & $+30.61$\\
      & \transA$_{\text{(b)}}$ & 38667.67 & 38637.1792(1) & \onlinecite{Riedle1994} & $+30.49$\\
      \benzC & \transA & 38725.98 & &\\
      \benzD & \transA & 38817.12 & 38785.935(10) & \onlinecite{Weber1991a} & $+31.19$\\
      \hline
      \benzAd & \transA & 38599.34 & $38561.9^*$ & \onlinecite{Boernsen1986} & $+37.4^*$\\
      & & & $38563.0^*$ & \onlinecite{Law1984} & $+36.3^*$\\
      & & 38602.24 & $38565.6^*$ & \onlinecite{Boernsen1986} & $+36.6^*$\\
      & & & $38566.7^*$ & \onlinecite{Law1984} & $+35.5^*$\\
      \benzBd & \transA & 38628.28 & & \\
      & & 38629.65(6) & & \\
      & & 38630.35(3) & & \\
      & & 38630.99(8) & & \\
      & & 38632.95 & & \\
      & & 38633.51 & & \\
      \benzCd & \transA & 38688.43 & & \\
      & & 38690.44(3) & & \\
      & & 38691.20 & & \\
      & & 38692.13 & & \\
      \benzDd & \transA & 38779.49 & $38741.7^*$ & \onlinecite{Law1984} & $+37.8^*$\\
      & & 38781.65(3) & & \\
      \hline
      \benzA & \transB & 39560.73 & 39529.636(3) & \onlinecite{Helm1996} & $+31.10$\\
      \benzAd & \transB & 39521.09(3) & & \\
      & & 39523.85 & & \\
      & & 39526.37 & & \\
      \hline
      \benzA--Ar & \transA & 38619.27 & 38585.071(8) & \onlinecite{Neusser1994} & $+34.20$\\
    \end{tabular}
  \end{ruledtabular}
\end{table}

\clearpage

\begin{table}
  \centering
  \caption{Solvation shifts of the electronic spectra (lowest allowed
  transition) of polyacenes in helium droplets.  \#ZPL refers to the
  number of ``zero-phonon'' lines observed in helium nanodroplets.
  The shift for anthracene was determined with only 12 helium atoms,
  and is expected to be different in helium droplets.}
  \label{tb:shifts}
  \begin{ruledtabular}
    \begin{tabular}{rlccc}
      & molecule & shift/\wn & \#ZPL & reference\\
      \hline
      \molec{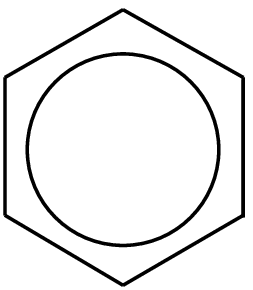} & benzene & $+30.31(2)$ & 1 & this work\\
      \molec{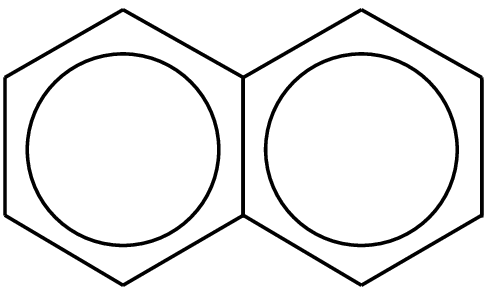} & naphthalene & $+15$ & 1 & \onlinecite{Lindinger1998}\\
      \molec{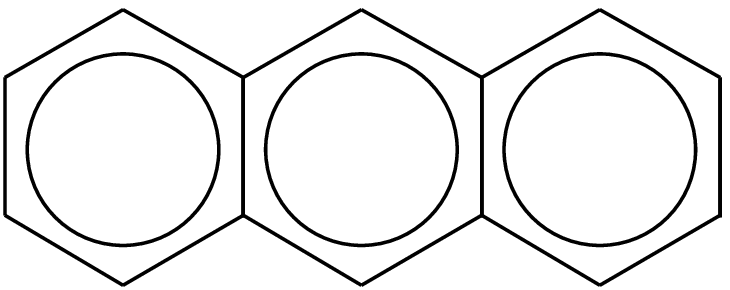} & anthracene & $(-12)$ & & \onlinecite{Even2001}\\
      \molec{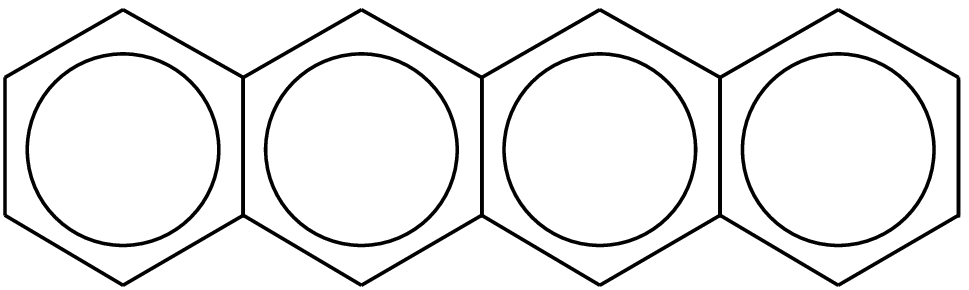} & tetracene & $-104.0(5)$ & 2 & \onlinecite{Hartmann2001}\\
      \molec{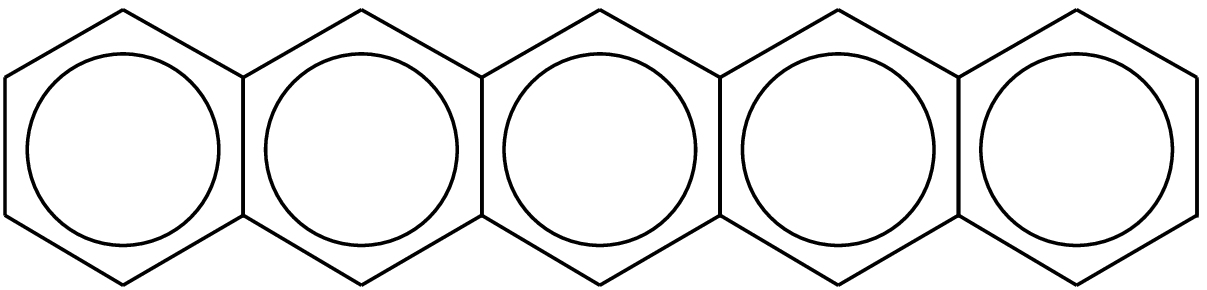} & pentacene & $-104.0(5)$ & 1 & \onlinecite{Hartmann2001}\\
    \end{tabular}
  \end{ruledtabular}
\end{table}

\clearpage

\begin{table}
  \centering
  \caption{Comparison of the electron distribution and polarizabilities in
  the ground (S$_0$) and first excited (S$_1$) states of \benzA,
  adapted from the CCSD calculations of
  Ref.~\onlinecite{Christiansen1999} (Sadlej+6 basis set).
  Experimentally, the quadrupole moment in the ground state is
  $-2.1(1)$\,e\AA$^2$ (Ref.~\onlinecite{Vrbancich1980}) or
  $-1.8(1)$\,e\AA$^2$ (Ref.~\onlinecite{Battaglia1981}).}
  \label{tb:moments}
  \begin{ruledtabular}
    \begin{tabular}{crrl}
      & S$_0$ & S$_1$ & \\
      \hline
      \multicolumn{4}{c}{Electron charge distribution}\\
      $\left< x^2 \right>=\left< y^2 \right>$ & 60.09 & 60.32 & \AA$^2$\\
      $\left< z^2 \right>$ & 8.43 & 8.48 & \AA$^2$\\
      \multicolumn{4}{c}{Quadrupole moment}\\
      $Q_{z z}$ & $-1.56$ & $-1.39$ & e\AA$^2$\\
      \multicolumn{4}{c}{Polarizability volume}\\
      $\alpha_{x x}=\alpha_{y y}$ & 11.96 & 12.30 & \AA$^3$\\
      $\alpha_{z z}$ & 6.65 & 7.74 & \AA$^3$\\
    \end{tabular}
  \end{ruledtabular}
\end{table}

\clearpage

\section*{Figure Captions:}

\begin{enumerate}[{FIG}.\ 1:]
\item Beam depletion spectra of \benzA, \benzB, \benzC, and \benzD\ 
  (left to right) in helium droplets. The peak positions are given in
  Table~\ref{tb:lines}.
\item Beam depletion spectra of the \transA\ (solid) and \transB\ 
  (dashed) lines of \benzA\ in helium droplets.  Wavenumbers are
  relative to the gas phase transitions (see Table~\ref{tb:lines}). No
  phonon wing is seen within 15\,\wn\ of the \transA\ transition. The
  dotted line is a fluorescence excitation spectrum of the \transA\ 
  transition, with the intensity (right-hand scale) referring to the
  estimated number of primary electrons in the photomultiplier tube
  per laser shot.
\item Panel A: comparison of the \benzA\ \transA\ droplet spectrum
  (solid line) and the gas phase spectrum at 0.38\,K (dashed line,
  from Ref.~\onlinecite{Okruss1999}; convoluted with a 0.2\,\wn\ FWHM
  Gaussian and shifted by $+30.31$\,\wn; stick spectrum in panel B).
  The low-temperature populations are mostly due to the conservation
  of nuclear spin, as the comparison to a spin-relaxed spectrum
  (dotted line; stick spectrum in panel C) shows. At zero temperature,
  only states with $J=K\le3$ are populated.
\item Fits (dashed lines) of the \benzA\ \transA\ beam depletion
  spectrum (solid lines) at 0.38\,K, using the rotational Hamiltonian
  of Ref.~\onlinecite{Okruss1999}. The rotational constants $B''$ and
  $B'$ are scaled by a numerical factor $\kappa_B$; $C''$ and $C'$
  (including $C_0'\zeta'$) are scaled by $\kappa_C$. The fitting
  procedure optimizes $\kappa_B$, $\kappa_C$, the width of the
  smoothing function (Gaussian), offset, amplitude, and blueshift. In
  the upper panel, room-temperature nuclear spin weights were assumed;
  in the lower panel, the nuclear spin weights were relaxed to their
  values at 0.38\,K. Both fits are equally good, yielding no
  information on the importance of nuclear spin relaxation. The stick
  spectra are magnified 10 times; they were convoluted with Gaussians
  of $\sim$0.4\,\wn\ FWHM to produce the spectral fits.
\item Beam depletion spectra of homo-dimers of four benzene
  isotopomers in helium nanodroplets.  Wavenumbers are relative to the
  monomer absorption lines as listed in Table~\ref{tb:lines} (in the
  case of \benzBd, relative to the average of the two lines). Solid
  lines are \transA\ spectra; the dashed line is an \transB\ spectrum.
\item Ratio of the intensity of the two \benzAd\ transitions as a
  function of argon pressure in the second pickup cell, with
  $\sim$\pt{8}{-4}\,mbar resulting in one argon pickup on average.
  The error bars denote $1\sigma$ intervals from least-squares fits of
  two Gaussians and a sloped baseline to the various spectra.
\item \transA\ overview beam depletion spectrum of \benzAd,
  \benzA--Ar, and \benzA. For this spectrum, a first pickup cell was
  filled with \benzA\ to a pressure that resulted in equal amounts of
  single and double pickups, and a second cell was optimized for
  pickup of a single argon atom (about \pt{8}{-4}\,mbar).  No further
  complexes with argon were detected.
\end{enumerate}

\clearpage

\begin{figure}
  \begin{center}
    \mygraph{monomerlines_sep}
    \caption{R.\ Schmied et al.}
    \label{fig:monomerlines}
  \end{center}
\end{figure}

\clearpage

\begin{figure}
  \begin{center}
    \mygraph{C6H6_monomer_A00}
    \caption{R.\ Schmied et al.}
    \label{fig:C6H6_monomer_A00}
  \end{center}
\end{figure}

\clearpage

\begin{figure}
  \begin{center}
    \mygraph{comparespectra_withsticks}
    \caption{R.\ Schmied et al.}
    \label{fig:comparespectra}
  \end{center}
\end{figure}

\clearpage

\begin{figure}
  \begin{center}
    \mygraph{stickspectrum}
    \caption{R.\ Schmied et al.}
    \label{fig:stickspectrum}
  \end{center}
\end{figure}

\clearpage

\begin{figure}
  \begin{center}
    \mygraph{dimerlines}
    \caption{R.\ Schmied et al.}
    \label{fig:dimerlines}
  \end{center}
\end{figure}

\clearpage

\begin{figure}
  \begin{center}
    \mygraph{lineratio}
    \caption{R.\ Schmied et al.}
    \label{fig:lineratio}
  \end{center}
\end{figure}

\clearpage

\begin{figure}
  \begin{center}
    \mygraph{overview}
    \caption{R.\ Schmied et al.}
    \label{fig:overview}
  \end{center}
\end{figure}

\end{document}